\documentclass{article}

\begin{document}
\begin{center}
{\Large\bf
Novel {\em coherent} quantum bit
using spatial quantization levels
 in semiconductor quantum~dot}\\

\medskip

L. Fedichkin, M. Yanchenko, and K.A. Valiev\\
\medskip
\small
{\em Institute of Physics and Technology\\
 Russian Academy of Sciences\\
34, Nakhimovsky pr.,\\
 Moscow, 117218, Russia}\\
E-mail:leonid@ftian.oivta.ru
\end{center}

\section{Introduction}
	Within last decade there has been achieved
large progress in theory of quantum computing.
Unfortunately the experimental realization
of practically valuable quantum computers has not been
performed mainly due to lack of scalable coherent controllable
two-level quantum systems.
Semiconductor nanostructures were considered as a candidates
to basic elements of quantum computer --- quantum bits (qubits).
In 1995 A.~Barenco, D.~Deutsch, A.~Ekert, and R.~Jozsa~\cite{Barenco}
for the first time offered to use as base states
("0" and "1") of qubits two bottom levels of spatial quantization of
single-electron quantum dots. To implement two-qubit operations
it was suggested to use electrical dipole interaction. Soon the same authors
(A.~Ekert and R.~Jozsa)~\cite{Ekert} evaluated coherence of such systems
and proved that ordinary quantum dots are too incoherent to
be quantum bits. The variants of such structures were investigated
by other authors in papers~\cite{Tanamoto,Sanders}.
Potential confining at least one of these quantum dots was assumed
asymmetrical. Distance between the bottom levels was about 10--100\,meV.

	Our proposal is to use mesoscopic structures with small (below 1\,meV)
separation between energies of two bottom states. The work frequency
of such structures, which is proportional to energy separation,
is surely to be decreased compared to early proposed ones.
But processes of spontaneous emission of photons and phonons
are proportional to polynom (cubic or higher degree) of energy of transition.
Therefore errors rate per one implemented quantum operation
is to be smaller compared to common structures.

\section{Structure and principles of operation}

	In the proposed structure we offer to use as a
qubit a quantum dot with a symmetrical potential profile, as shown in
Figs.~1, 2.
The presence of two minima of potential separated by a
thick barrier is essential. In a quantum
dot there is one electron. The presence of the second electron is
energetically unprofitable because of Coulomb interelectron repulsion.
Although the qubit proposed  can be made in principle
of any semiconductor for technological reasons we focus our
attention on GaAs/AlGaAs structures
because this material proved to reliable basis for observation of
various coherent quantum effects.
In GaAs quantum dots at distance between
minima $r=10$\,nm (see Fig.~3) the Coulomb energy is about
$e^2/\kappa r = 11$\,meV, that allows to exclude a spontaneous charging
of a dot by the second electron.

To find work frequencies and other working parameters of an offered qubit,
two-dimensional Schr\"odinger equation for an electron in a quantum dot
in GaAs with model potential $V$ (shown on Fig.~2) was solved numerically:

\begin{equation} \label{V}
V = \frac{{m\omega ^2 (x^2  + y^2 )}}{2} +
V_B \exp{\left[ - \frac{{x^2 }}{{\left( {wl} \right)^2 }}\right]},
\end{equation}
where $m = 0.065m_e$; $l = 20$\,nm; $V_B = 1.5 \cdot 10^{-19}$\,J;
$w = 0.08 \div 0.34$ ($r = 11 \div 34$\,nm).

The obtained wavefunctions of four bottom states of an electron for this
potential are shown in Figs.~4--7. Two bottom energy levels of an electron in
such point correspond to wavefunctions $\Psi_1$ (Fig.~4) and $\Psi_2$
(Fig.~5). For logic "0" and "1" it is convenient to take
the normalized sum and difference
of $\Psi_1$ and $\Psi_2$ rater than
states with the certain energy:

\begin{equation}
  \left| 0 \right\rangle  = \frac{1}
{\sqrt 2 }\left( \Psi _1 + \Psi _2  \right),\qquad
  \left| 1 \right\rangle  = \frac{1}
{\sqrt 2 }\left( \Psi _1 - \Psi _2  \right).
\end{equation}

The given states correspond to almost complete localization of an electron in one of
minima of potential, as shown in a Figs.~8, 9. It allows to implement write of
the input data and reading of results by methods of single-electronics with
the help of the read-out gates, as shown in a Fig. 1. The
central gate (control gate) serves for downturn of a potential barrier
while implementing quantum unitary transformations, as will be shown below.

\subsection{Implementation of one-qubit unitary operations}

In the beginning qubit is not affected by operations and it is in the
state

\begin{equation} \label{psi0}
\Psi (0)  = c_0 \left| 0 \right\rangle  + c_1 \left| 1 \right\rangle.
\end{equation}
Electron wavefunction evolve according to Schr\"odinger equation
\begin{equation}
i\hbar \frac{d\Psi}{dt} = H\Psi ,
\end{equation}
where Hamiltonian
\begin{equation}
H =  - \frac{{\hbar ^2 }}{{2m}}\Delta  + V,
\end{equation}
where $V$ is given by (\ref{V}).

Different voltages on the control gate correspond to different barrier heights $V_B$.
Therefore energies of the ground and first excited state also depend on $V_B$.
$\bigl(\hbar\omega_{1,2}=\hbar\omega_{1,2}\left(V_B\right)\bigr)$.
When barrier is high, the energy levels of two bottom states practically merge
so two bottom levels evolve with the common frequency~$\omega$:

\begin{equation}
\Psi (t) = e^{-i\omega t} \left( {c_0 \left| 0 \right\rangle
                               + c_1 \left| 1 \right\rangle } \right).
\end{equation}

The downturn of a barrier results in an inequality of frequencies ($\Delta
\omega = \omega_2 -\omega_1 >0$) and to periodic rotation of vector of state
of qubit in basis
{($\left| 0 \right\rangle,\left| 1 \right\rangle$)}.

\begin{equation}
\Psi (t) = \frac{1}{{\sqrt 2 }}\left( {c_0  + c_1 } \right)\Psi _1
  e^{-i\omega _1 t}  + \frac{1}{{\sqrt 2 }}\left( {c_0  - c_1 } \right)
  \Psi _2 e^{-i\omega _2 t} ,
\end{equation}
that is

\begin{eqnarray}
  \Psi (t) &=& e^{-i(\omega _1  + \omega _2 )t/2}
    \bigl[\left( {c_0 \cos \Delta \omega t/2 + i\,c_1 \sin \Delta \omega t/2} \right)
       \left| 0 \right\rangle\bigr.  +\\
  & &+ \bigl.\left( {c_1 \cos \Delta \omega t/2 + i\,c_0 \sin \Delta \omega t/2} \right)
       \left| 1 \right\rangle\bigr].\nonumber
\end{eqnarray}

Having applied a pulse of a positive voltage of the certain duration
$t_{NOT}$, equal to $\pi/\Delta\omega$, we
(an insignificant common phase factor
$\exp \left[  - i\pi {\omega _1 }/
{(\omega _2  - \omega _1)} \right]$ can be neglected)
shall transfer a state of a qubit~(\ref{psi0})
into a state (as shown in Fig.~10):

\begin{equation}
{\rm NOT}(\Psi _0 ) =
c_1 \left| 0 \right\rangle  + c_0 \left| 1 \right\rangle.
\end{equation}

So, with the help of the given procedure it is possible to exchange amplitudes
at "0" and "1", that is to carry out unitary NOT operation. Changing the
duration of a pulse it is possible to carry out rotation of qubit state
to any required angle.

\subsection{CNOT gate implementation}

For construction of the universal quantum computer it is also necessary to be
able to realize at least one nontrivial (not decomposable into a sequence of
one-qubit gates and permutations) two-qubit operation.
We consider realization of CNOT operation.
(Depending on state of control qubit, on target qubit should be carried
out either operation of identity or NOT). For implementation of the
CNOT gate between
the neighbour qubits the Coulomb interaction is used. We arrange two qubits
as shown in a Fig.~3. To exclude exchange effects it is supposed
that the qubits are separated by completely opaque barrier.
Then Hamiltonian of electron in target qubit becomes
\begin{equation}
H_t =  - \frac{{\hbar ^2 }}{{2m}}\Delta  + V+V_C,
\end{equation}
where Coulomb potential due to electron in control qubit
\begin{equation}
V_C \left( {x,y} \right) = \int{\!\!\!\int {du\,dv\,\frac{{\left| {\Psi_C \left( {u,v} \right)} \right|^2 e^2 }}{{\kappa\sqrt {\left( {x + v} \right)^2  + \left( {y + R - u} \right)^2 } }}} },
\end{equation}
where $\Psi_C\left( {u,v}\right)$ is a wavefunction of electron in control qubit,
$u,v$ are coordinates in coordinate system  of control qubit, $e$ is the
electron charge and $R$ is a separation between qubits' centers. Height of barrier dividing the right (target) qubit depends on state of the left (control) qubit.
The addition to the height of barrier of the target qubit can be roughly estimated in the following way: when control qubit is in state $\left| 0 \right\rangle$ or $\left| 1 \right\rangle$
we can assume that electron is a point charge located in electron density maximum.
Then the Coulomb potential in the barrier dividing target qubit will be
\begin{equation}
V_C \left( {0,y} \right) = {{e^2 } \mathord{\left/
 {\vphantom {{e^2 } {\left( {y + R + s{r \mathord{\left/
 {\vphantom {r 2}} \right.
 \kern-\nulldelimiterspace} 2}} \right)}}} \right.
 \kern-\nulldelimiterspace} {\kappa\left( {y + R + s\,{r \mathord{\left/
 {\vphantom {r 2}} \right.
 \kern-\nulldelimiterspace} 2}} \right)}},
\end{equation}
where $s$ corresponds to state of control qubit: $s=1$, when control qubit
is in state $\left| 0 \right\rangle$, and $s=-1$, when control qubit
is in state $\left| 1 \right\rangle$.

In such structure CNOT gate can be implemented as follows. Having slightly opened
the barrier in a target qubit with the help of its control gate it is possible
to achieve application to the target qubit the operation of identity,
when the control qubit is in a state $\left| 0 \right\rangle$ and operation
NOT, when the control qubit is in a state $\left| 1 \right\rangle$, that is
operation CNOT. Consider a case, when
the control qubit is in one of base states. Because of Coulomb influence of a
control qubit on height of a barrier in a target  qubit the duration of the
NOT operation in these cases will differ ($t_{NOT0} \neq t_{NOT1}$).
Duration of a pulse on a gate of a target  qubit we choose in the
following way:
\begin{equation}
t_{CNOT} = t_{NOT0}\, t_{NOT1} /(t_{NOT1} - t_{NOT0}).
\end{equation}
Varying amplitude of a pulse we shall achieve, that the ratio
\begin{equation}
t_{NOT1} /2(t_{NOT1} - t_{NOT0})
\end{equation}
will be an integer
(it is the large value, as in our case interaction is weak, so
by small change of amplitude we attain the nearest integer value)
\begin{equation}
t_{NOT1} /2(t_{NOT1} - t_{NOT0})=N.
\end{equation}
If the control qubit is in state "0", then the action of a pulse is equivalent
to consecutive application of {\em even} $(2N)$ of number of operations NOT to the
second qubit, that is operation of identity. If the control qubit is in state
"1", then the action of a pulse is equivalent to consecutive application of
{\em odd}
$(2N-1)$ number of operations NOT to the second qubit, that is NOT operation.

While modelling the interaction was calculated directly
by the Coulomb's law
(in all cells of a two-dimensional grid the field created by a partial charge
of a control qubit form all cells) was calculated.
Exchange effects were neglected. The dependencies of minimal durations of
operations NOT and CNOT depending on geometrical parameters are shown in
Fig.~11, 12.

\section{Numerical modeling}

For calculation the method of simple iterations was used.
Potential should be symmetrical and have 2 minima, that means presence of a
barrier between these minima. Similar potential for a two-electronic quantum dot
was used in \cite{DiVincenzo}: he has offered to use polynomial potential of
the 4-th order like
\begin{equation}
V\left( {x,y} \right) = \frac{{m\omega ^2 }}{2}\left( {\frac{1}{{4a^2 }}
\left( {x^2  - a^2 } \right)^2  + y^2 } \right)
\end{equation}
with minima in points~$\pm a$.

In this work as potential of a quantum dot with the built-in tunnel barrier
was used potential~(\ref{V}).
Such potential can be varied by changing the
following parameters:
$l$ (characteristic size of a qubit ($\sim 20$\,nm)), $V_B$
(height of a barrier ($\sim 1$\,eV)) and $w$ (the width of a barrier in relation to
$l$ ($\sim 0.1$)). Varying values of these parameters, it is possible to
tune the characteristics of structure.

Consider an electron strongly limited in a direction, perpendicular to
a surface of a heterojunction. Thus distances between levels of spatial
quantization (2D electron gas subbands) in this direction is about 100\,meV.
At low temperatures and weak influences the electron remains in the
bottom subband. Thus its motion can be considered as two-dimensional,
and the wavefunction of an electron is factorized into
\begin{equation}
\Phi \left( {x,y,z} \right) = \Psi \left( {x,y} \right)g\left( z \right).
\end{equation}
For wavefunction $ \Psi \left( {x,y} \right)$ we have solved 2D Schr\"odinger
equation with the potential $V$ using simple iterations method with
orthogonalization on each iteration. We have modelled an area $60 \times 40\,
\mathrm{nm}^2$ using mesh with the same step in both directions 0.5\ nm. We have
iterated 4 bottom state wavefunctions controlling their orthogonality. The
mean number of iterations was about 10000. To find energies of states of an
electron, the mean value of an energy for a wavefunction $\Psi$  under the
formula
$\left\langle \Psi  \right|H\left| \Psi  \right\rangle $,
where $H$ is the operator of Hamilton with real potential was calculated.
After that the program show us the diagram of the bottom part of a spectrum
of system, and also frequency of transitions from the exited energy levels on
basic. These frequencies were also used in research of processes of decoherence in
structure. Besides half of time of transition from the first exited level on
basic and back also was accepted for the minimal time of operation of the NOT
gate.

For the nontrivial two-qubit CNOT gate it is necessary to include interaction
between control and target  qubits. In studied structure the electrical
(mostly dipole) interaction is used. For this purpose we arrange a target
qubit about an end face of a control qubit, having turned it on 90 degrees in
a plane of structure. At such arrangement a state of an electron in a control
qubit (the electron is farther or closer to the middle of a target  qubit)
effectively reduces or increases height of a barrier in a target  qubit,
that changes time of operation of the NOT gate. As the points of localization
of an electron in a target  qubit are located symmetrically in relation to
the control qubit, so the change of a state of a target  qubit does not
render influence in the first order on a state of a control qubit.

For calculation of interaction it is necessary to calculate potential created
by an electron of a control qubit in the field of a target  qubit. For this
purpose it is possible to take advantage of two methods: solution of a
Poisson equation or calculation of an electrostatic field using Coulomb's
law. The second method was used. For this purpose it was supposed, that in
each node of a mesh in the area of a control qubit there is a point charge,
which value is equal to value of a charge of an electron multiplied on a value
of function of spatial distribution of an electron at this node of a mesh. As
function of distribution it is natural to take a square of the module of a
wavefunction. As the function of distribution is normalized, this operation
simply represents an electron by system of point charges with a cumulative
charge equal to a charge of an electron.

The durations of the NOT gate for two extreme states of an electron
in a control qubit were obtained: when the electron wholly is in most distant
from a target  qubit and when the electron wholly is near to a target
qubit. Knowing these times, it is possible to calculate duration of operation of
the CNOT gate.

\section{Coherence of structure}

Significant difficulty interfering creation of the large scale quantum
computer is the problem of decoherence of a quantum state because of
interaction with an environment bringing in errors.
It is proved
\cite{Kitaev}, that if the decoherence occurs slowly enough, that, in
particular, means, that at calculation occurs no more than $\delta$ failures
for one computing step (by various estimations $\delta$ should lie in a
range from $10^{-2}$ to $10^{-5}$), with the help of special
error correcting algorithms and codes
demanding polynomial increase of computing expenses,
modeling functioning of ideal (coherent) quantum computer
and steady implementation of any quantum algorithm is possible.

While calculating the decoherence in offered structure the low-temperature
limit $\left(T \rightarrow 0\right)$ was considered. The given approximation
is justified, as the modern cryogenic engineering allows to carry out
functioning nanoelectronic structures at temperatures down to several
millikelvins, that is sufficiently lower than
the distance between bottom and first excited levels.
The case of high temperatures represents only academic
interest because of inevitable fastest decoherence and impossibility of correct
work of the quantum computer.
However, in solid-state structures even at
absolute zero of temperature
the processes of decoherence owing to spontaneous
emission of photons or acoustic phonons with transition of an electron
from excited to the basic level are possible.
These processes will determine
the degree of coherence in our structure.
We separately investigated spontaneous
emission of  photon, of deformation acoustic
phonon and of piezoelectric acoustic phonon.

\subsection{Emission of photon}

Consider process of decoherence through emission of photon. The
transition from first exited on the basic level is dipole. The probability of
dipole transition is given by the known formula~\cite{Landau4}
\begin{equation}
W_{Ph}  = \frac{{4\omega ^3 }}{{3\hbar c^3 }}\left| {\bf d} \right|^2,
\end{equation}
where the dipole moment
\begin{equation}
{\bf d} = \int {d^2 r\,\Phi '\left( {\bf r} \right)^* e{\bf r}\,\Phi
          \left( {\bf r} \right)} .
\end{equation}
From the symmetry of wavefunctions it follows that of a component $d_y$ will
be zero, and $x$-component can be calculated through integral not on all
space, but only on half-space $x>0$:
\begin{equation}
d_x  = 2\int\limits_{x > 0} {dx\,dy\,\Phi '\left( {\bf r} \right)^* ex\,\Phi
       \left( {\bf r} \right)} .
\end{equation}
As the integrand at large $r$ exponentially decreases with increasing of $x$,
integral can approximately be evaluated, having reduced integration on
half-space to integration on area $0<x<x_{max}$.
To estimate this integral we can replace $x$ in an
integrand with its maximal value,
i.e.
\begin{equation}
d_x  \leq 2ex_{\max } \!\!\!\!\!\!\!\!\int\limits_{0 < x < x_{\max } }
 {\!\!\!\!\!\!\!\!dx\,dy\,\Phi '
     \left( {\bf r} \right)^* \Phi \left( {\bf r} \right)} ,
\end{equation}
and as the wavefunctions on area of integration practically coincide,
integral will be equal to $1/2$, as the integration is made on half-space, and
integral on all space from a condition of a normalization of wavefunctions is
equal to 1. In our case of wavefunctions have on $x>0$ maxima in some point
$r$, behind which exponentially fall down, and as $x_max$ it is enough to choose
$2r$. Then $d_x \leq 2er$, hence estimation for probability of spontaneous emission
of photon to look like

\begin{equation}
W_{Ph}  \leq \frac{{16\varepsilon_{10}^3 e^2 r^2 }}{{3\hbar ^4 c^3 }}.
\end{equation}

\subsection{Emission of acoustic phonon}

The probability of  relaxation from the excited state to the ground one,
by emission of acoustic phonon, is calculated by formula:
\begin{equation}
w_{10}  = \frac{{2\pi }}{\hbar }\left| M \right|^2 \delta
          \left( {\varepsilon _{10}  - \hbar sq} \right),
\end{equation}
where $M = \left\langle f \right|T\left| i \right\rangle $ ($\left| i
 \right\rangle$ is the initial state
(the electron is in the excited state with
 an energy $\varepsilon_1$),
$\left| f \right\rangle$ is the final state (the
electron is in the base state with an energy $\varepsilon_0$,
$\varepsilon_{10}  \equiv \varepsilon _1  - \varepsilon _0,$
the phonon carried away an energy $\hbar sq$),
$T$ is the transition operator) is the
matrix element of transition appropriate to emission of  phonon, $s$ is the
speed of sound ($5.2\cdot 10^3$ m/s in GaAs), ${\bf q}$ is a wave vector of
the phonon.

Disturbance created by one phonon $s{\bf q}$~\cite{Levinson}:
\begin{equation} \label{Vsq}
V_{s{\bf q}}  = \frac{1}{{L^{{3 \mathord{\left/
 {\vphantom {3 2}} \right.
 \kern-\nulldelimiterspace} 2}} }}\left[ {\frac{{\hbar a_0^3 }}
          {{2M_0 \omega _{s{\bf q}} }}} \right]^{{1 \mathord{\left/
 {\vphantom {1 2}} \right.
 \kern-\nulldelimiterspace} 2}} e^{i{\bf qr}} v_{s{\bf q}}
          \left( {\bf r} \right),
\end{equation}
and
\begin{equation}
\label{vsq}
v_{s{\bf q}} \left( {\bf r} \right) = \sum\limits_{{\bf a}\alpha }
      {V_{{\bf a}\alpha } \left( {\bf r} \right){\bf d}_{s{\bf q}}^\alpha  }
      e^{i{\bf q}\left( {{\bf a} - {\bf r}} \right)} ,
\end{equation}
where $M_0$ is the mass of an elementary cell, $a_0^3$ is the volume of an
elementary cell, $d$ is a unitless vector of  polarization orthonormalized
by the condition
\begin{equation}
\sum\limits_\alpha  {M_\alpha  \left( {{\bf d}_{s{\bf q}}^\alpha  } \right)^*
     {\bf d}_{s'{\bf q'}}^\alpha  }  = M_0 \delta _{s{\bf q},s'{\bf q'}} ,
\end{equation}
where $M_\alpha$ is a mass of atom $\alpha$. From a normalization follows,
that for acoustic phonons at ${\bf q} \to 0$ all atoms in a cell are displaced
equally:
\begin{equation}
{\bf d}_{s{\bf q}}^\alpha   = {\bf d}_{s{\bf q}}^{} ,
\qquad\left| {{\bf d}_{s{\bf q}}^{} } \right| = 1.
\end{equation}

It is convenient to define a total probability of emission of  phonon, i.e.
probability to emit though any phonon, considering, that the crystal is
limited to a normalizing volume $L^3$, and the allowable values of a pulse are
quasidiscrete. Then the probability of transition is probability that in unit
of time electron will make a transition from one quasidiscrete state in
another; the dimensional representation of it is $s^{-1}$. The certain thus
probability depends on a normalizing volume. Probability of  transition of an
electron from the given state
\begin{equation}
W = \sum\limits_{\bf q} {w_{\bf q} } .
\end{equation}
As the allowable values of q are located very richly, on distance $2\hbar/L$,
it is possible to pass from a summation on ${\bf q}$ to integration by a rule

\begin{equation}
\sum\limits_{\bf q} {\left(  \ldots  \right)}  = L^3 \int {\frac{{d^3 q}}
  {{\left( {2\pi } \right)^3 }}\left(  \ldots  \right)} .
\end{equation}
Then
\begin{equation} \label{W}
W = L^3 \int {\frac{{d^3 q}}{{\left( {2\pi } \right)^3 }}w_{10} } .
\end{equation}

\subsection{Emission of piezoelectric acoustic phonon}

In~1961 A.R.~Hutson considered interaction
of electrons and acoustic waves due to piezoelectric effect~\cite{Hutson}.
The matrix element of interaction of an electron with an electromagnetic field
is given by the formula
\begin{equation}
M_{PA}  = \int {d^3 r\,\Phi '\left( r \right)^* e\varphi \left( r \right)\Phi
           \left( r \right)} ,
\end{equation}
where $\Phi \left( r \right)$ is wavefunction
of an initial state of electron $\left| i \right\rangle $, $\Phi '
\left( r \right)$ - of the final state $\left| f \right\rangle $ ,
$\varphi \left( r \right)$  is the macrofield created by a phonon.
It can be found from the Poisson equation
\begin{equation} \label{poisson}
\nabla ^2 \varphi  = 4\pi\,\mathrm{div}{\bf P},
\end{equation}
where ${\bf P}$ is a dipole moment of unit of volume arisen at deformation of
lattice. Use of the Poisson equation instead of
complete system of the equations of the Maxwell corresponds to the
indefinitely large speed of light $c$. This assumption is justified, as $c$ is
much greater of phase speed $\omega / q$ of phonons, that are interesting for
us.

At homogeneous acoustic deformation
\begin{equation} \label{P_j}
P_j  = \beta _{jkl} u_{kl} ,
\end{equation}
where $\beta$ - piezoelectric constants. Having substituted (\ref{P_j}) into
(\ref{poisson}), it is possible to find a field $\varphi$. If we are interested
in a field $\varphi _{s{\bf q}}$, created by single phonon $s{\bf q}$, from
the Poisson equation we have
\begin{equation}
\varphi _{s{\bf q}}  =  - i({{4\pi } \mathord{\left/
 {\vphantom {{4\pi } {q^2 }}} \right.
 \kern-\nulldelimiterspace} {q^2 }}){\bf qP}_{s{\bf q}} ,
\end{equation}
where ${\bf P}_{s{\bf q}}$ is a polarization created by one phonon $s{\bf q}$.

As a result for acoustic phonons it turns out \cite{Levinson}
\begin{equation}
\varphi \left( {{\bf r},t} \right)_{s{\bf q}}  = \frac{1}{{L^{{3
 \mathord{\left/ {\vphantom {3 2}} \right.
 \kern-\nulldelimiterspace} 2}} }}\left[ {\frac{{\hbar a_0^3 }}{{2M_0 \omega
 _{s{\bf q}} }}} \right]^{{1 \mathord{\left/
 {\vphantom {1 2}} \right.
 \kern-\nulldelimiterspace} 2}} e^{i{\bf qr} - i\omega _{s{\bf q}} l}
 \beta _{s{\bf q}}  + c.c.,
\end{equation}
where the effective piezoelectric constant of  wave $s{\bf q}$ is entered:
\begin{equation}
\beta _{s{\bf q}}  = 4\pi e_i e_k \beta _{ikj} d_{s{\bf q}}^j ,
\qquad
e = {{\bf q} \mathord{\left/
 {\vphantom {{\bf q} {q;}}} \right.
 \kern-\nulldelimiterspace} {q;}}
\end{equation}
it depends only on a direction of phonon propagation and from a polarization.
Thus, the probability to emit a piezoelectric acoustic phonon
\begin{equation} \label{w_10}
w_{10}  = \frac{{\pi \hbar }}{{\rho \varepsilon _{10} L^3 }}
   \left( {e\beta _{s{\bf q}} } \right)^2 \left| {I\left( {\bf q} \right)}
 \right|^2 \delta \left( {\varepsilon _{10}  - \hbar sq} \right),
\end{equation}
where
\begin{equation}
I\left( {\bf q} \right) = \int {d^2 r\,\Phi '\left( {\bf r} \right)^*
 \Phi \left( {\bf r} \right)e^{i\left( {q_x x + q_y y} \right)} } .
\end{equation}

With the account of (\ref{w_10}) for a total probability we finally have
\begin{equation}
W_{PA}  = \frac{{\pi \hbar }}{{\left( {2\pi } \right)^3 \rho \varepsilon _{10}
 }}\int {d^3 q\left( {e\beta _{s{\bf q}} } \right)^2 \left| {I\left( {\bf q}
\right)} \right|^2 \delta \left( {\varepsilon _{10}  - \hbar sq} \right)} .
\end{equation}

For calculation of probability it is convenient to pass to spherical
coordinates. This transition was carried out by a rule

\begin{equation} \label{spher}
\left\{ \begin{array}{l}
 q_x  = q\cos \theta \cos \varphi , \\
 q_y  = q\cos \theta \sin \varphi , \\
 q_z  = q\sin \varphi . \\
 \end{array} \right.
\end{equation}

In spherical coordinates the expression for calculation of probability will
be
\begin{equation}
W_{PA}  = \frac{{\pi \hbar }}{{\left( {2\pi } \right)^3 \rho \varepsilon _{10}
 }}\int {d\theta\, d\varphi\, dq\,q^2 \cos \theta \left( {e\beta _{s{\bf q}} }
 \right)^2 \left| {I\left( {\bf q} \right)} \right|^2 \delta
 \left( {\varepsilon _{10}  - \hbar sq} \right)}
\end{equation}
We get rid of delta-function under integral, having made integration on $dq$:
\begin{equation}
W_{PA}  = \frac{{\pi \hbar }}{{\left( {2\pi } \right)^3 \rho \varepsilon _{10}
 }}\left( {\frac{{\varepsilon _{10} }}{{\hbar s}}} \right)^2 \frac{1}{{\hbar
s}}\int {d\theta\, d\varphi\, \cos \theta \left( {e\beta _{s{\bf q}} } \right)^2
\left| {I\left( {\bf q} \right)} \right|^2 } .
\end{equation}
and finally we get
\begin{equation}
W_{PA}  = \frac{{\varepsilon _{10} }}{{8\pi ^2 \rho \hbar ^2 s^3 }}\int
 {d\theta\, d\varphi\, \cos \theta \left( {e\beta _{s{\bf q}} } \right)^2
 \left| {I\left( {\bf q} \right)} \right|^2 } .
\end{equation}

Vector of displacement of atoms is
\begin{equation}
{\bf d} = \left( {\begin{array}{*{20}c}
   a\\
   b\\
   c\\
\end{array}} \right) .
\end{equation}
We consider cubic crystals of symmetry classes $O_h$ and
$T_d$. In crystals $O_h$ with the center of inversion (for example, in Si, as
it is possible to show from reasons of symmetry, $\beta = 0$, i.e. the
homogeneous deformation does not create macrofields. In crystals $T_d$ without
the center of inversion (for example, GaAs) the tensor $\beta_{ikj}$ has only
those components, in which all three indexes $i, k, j$ are various, and all
these components are equal. Therefore

\begin{equation}
\beta _{\bf q}  = \frac{{4\pi \bar \beta }}{{q^2 }}
                  \left( {q_x q_y c + q_y q_z a + q_z q_x b} \right),
\end{equation}
where $\bar \beta$ is constant ($e_{14} / \kappa_0$, where $e_{14}$ is sole
piezoelectric constant of  cubic crystal ($0.16\, \mathrm{C} / \mathrm{m}^2$ for GaAs
\cite{Levinson,Ridley}), $\kappa_0$ - dielectric permeability (12.8 for GaAs
\cite{Levinson,Ridley})).
In next sections we will fulfil calculations of
probability of emission of  piezoelectric acoustic
phonon for GaAs.

\subsubsection{Transverse phonons}

The phonon with a transverse polarization propagated in any direction,
can be decomposed on basis consisting of two polarizations,
perpendicular to each other. Choose the first direction of
polarization not causing displacement of atoms in direction $z$.
The vector of displacement of atoms for the first direction of  polarization also should be
perpendicular to the wave vector $\bf q$ and to be normalized:
\begin{equation}
{\bf d}_{T1}  = \frac{1}{{\sqrt {q_x^2  + q_y^2 } }}
\left( {\begin{array}{*{20}c}
   {q_y }  \\
   { - q_x }  \\
   0  \\
\end{array}} \right).
\end{equation}
For such wave an effective piezoelectric constant
\begin{eqnarray}
  \beta _{T1{\bf q}} &=& \frac{{4\pi \bar \beta }}{{q^2 }}\left( { - q^2 \sqrt
 {\cos ^2 \theta } \cos 2\varphi \sin \theta } \right) =
\\&&   =  - 4\pi \bar \beta \sqrt {\cos ^2 \theta } \cos 2\varphi \sin \theta ,
\nonumber
\end{eqnarray}
and under integral there will be an expression

\begin{equation}
\left( {e\beta _{T1{\bf q}} } \right)^2  = \left( {4\pi e\bar \beta }
        \right)^2 \cos ^2 \theta \cos ^2 2\varphi \sin ^2 \theta .
\end{equation}

The vector of displacement for the second direction of  polarization should
be perpendicular to the wave vector $\bf q$, to the vector $\bf d$ and to be
normalized:
\begin{equation}
{\bf d}_{T2}  = \left( {\begin{array}{*{20}c}
   {\frac{{q_x q_z }}{{q_y q}}\sqrt {\frac{{q_y^2 }}{{q_x^2  + q_y^2 }}} }  \\
   {\frac{{q_z }}{q}\sqrt {\frac{{q_y^2 }}{{q_x^2  + q_y^2 }}} }  \\
   { - \frac{{q_y }}{q}\sqrt {\frac{{q_x^2  + q_y^2 }}{{q_y^2 }}} }  \\
\end{array}} \right).
\end{equation}
Effective piezoelectric constant for this direction of  polarization
\begin{eqnarray}
  \beta _{T2{\bf q}}  &=& \frac{{4\pi \bar \beta }}{{q^2 }}\left( { -
 \frac{1}{4}q^2 \left( {\cos \theta  + 3\cos 3\theta } \right)\cos \varphi
 \sqrt {\sin ^2 \varphi } } \right) =\\
 & & =  - \pi \bar \beta \left( {\cos \theta  + 3\cos 3\theta } \right)
 \cos \varphi \sqrt {\sin ^2 \varphi } ,\nonumber
\end{eqnarray}
and under integral there will be an expression
\begin{equation}
\left( {e\beta _{T2{\bf q}} } \right)^2  = \left( {\pi e\bar \beta } \right)^2
 \left( {\cos \theta  + 3\cos 3\theta } \right)^2 \cos ^2 \varphi
 \sin ^2 \varphi .
\end{equation}

Summarizing on polarizations of phonons, thus, in the integrand we
obtain an expression
\begin{eqnarray}
  \left( {e\beta _{T{\bf q}} } \right)^2 & \equiv& \left( {e\beta _{T1{\bf q}} }
 \right)^2  + \left( {e\beta _{T2{\bf q}} } \right)^2  = \left( {\frac{{\pi
   e\bar \beta }}{2}} \right)^2 \cos ^2 \theta  \times
\\ & &
   \times \left( {4\left( {7 - 9\cos 2\theta } \right)\cos 4\varphi \cos ^2
 \theta  - 28\cos 2\theta  + 9\cos 4\theta  + 27} \right).\nonumber
\end{eqnarray}

Finally, the total probability of emission of  phonon with a transverse
polarization is given by the following expression:

\begin{eqnarray}
  W_{TPA} & = &\frac{{\varepsilon _{10} \left( {e\bar \beta } \right)^2 }}
{{32\rho \hbar ^2 s^3 }}\int {d\theta\, d\varphi } \left| {I\left( {\varepsilon
 _{10} \frac{{\cos \theta }}{{\hbar s}},\varphi } \right)} \right|^2 \cos ^3
\theta  \times
\\ & &
   \times \left( {4\left( {7 - 9\cos 2\theta } \right)\cos 4\varphi \cos ^2
\theta  - 28\cos 2\theta  + 9\cos 4\theta  + 27} \right).\nonumber
\end{eqnarray}

\subsubsection{Longitudinal phonons}

The phonon with a longitudinal polarization propagated in some direction, will
cause displacement of atoms in the same direction:
\begin{equation}
{\bf d}_L  = \frac{1}{q}
\left( {\begin{array}{*{20}c}
   {q_x }  \\
   {q_y }  \\
   {q_z }  \\
\end{array}}
\right) .
\end{equation}
Effective piezoelectric constant for such vector
\begin{equation}
\beta _{L{\bf q}}  = \frac{{4\pi \bar \beta }}{{q^3 }}3q_x q_y q_z ,
\end{equation}
and the integral expression will get
\begin{equation}
  \left( {e\beta _{L{\bf q}} } \right)^2  = \frac{{\left( {12\pi e\bar \beta }
 \right)^2 }}{{q^6 }}\left( {q_x q_y q_z } \right)^2  =
   = \left( {12\pi e\bar \beta } \right)^2 \cos ^4 \theta \sin ^2 \theta
 \cos ^2 \varphi \sin ^2 \varphi .
 \end{equation}

Finally, the total probability of emission of  phonon with a longitudinal
polarization is given by the following expression:
\begin{equation}
W_{LPA}  = \frac{{18\hbar \left( {e\bar \beta } \right)^2 \varepsilon _{10} }}
{{\rho \left( {\hbar s} \right)^3 }}\int {d\theta\, d\varphi \left| {I\left(
 {\varepsilon _{10} \frac{{\cos \theta }}{{\hbar s}},\varphi } \right)}
\right|^2 \cos ^4 \theta \sin ^2 \theta \cos ^2 \varphi \sin ^2 \varphi}
\end{equation}

\subsubsection{Emission of deformation acoustic phonon}

The matrix element of interaction of an electron with a deformation
field introduced in~1950 by J.~Bardeen and W.~Shockley~\cite{Bardeen} is given by
\begin{equation}
M_{DA}  = \int {d^3 r\,\Phi '} \left( {\bf r} \right)^* w\left( {\bf r}
 \right)\Phi \left( {\bf r} \right) ,
\end{equation}
where value $w(r)$ is called deformation potential. The homogeneous
deformation of crystal due to long-wave acoustic phonons is described by a
tensor of deformation $u_{ij}$. Therefore in the lowest order on displacement
of atoms deformation potential of acoustic phonons can be expand into a
series of $u_{ij}$ and to write down
\begin{equation} \label{w_def}
w = \Xi ^{ij} u_{ij}
\end{equation}
(the summation on repeating indexes is performed).
Values $\Xi^{ij}$ (of the units of energy)
are referred as constants of deformation potential. As the tensor
$u_{ij}$ is symmetrical, the same property has the tensor of constants of
deformation potential $\Xi^{ij}$. The number of independent constants is
determined by symmetry of  Brillouin zone in that point $\bf k$, in which the
influence of deformation on a spectrum is studied.

The elementary case takes place in a cubic crystal, when the extremum of the
band is located in a point ${\bf k} = 0$. The symmetrical tensor of the
second rank in this case is reduced to one constant
\begin{equation}
\Xi ^{ij}  = \delta _{ij} \Xi .
\end{equation}
Then
\begin{equation}
w = \Xi u, \qquad u = u_{11}  + u_{22}  + u_{33},
\end{equation}
where $u$ - relative change of volume as a result of deformation. In this case
deformation of shift which is not changing of volume, does not result in
occurrence of deformation potential. From (\ref{Vsq}) and (\ref{vsq}) one
obtain
\begin{equation} \label{u_ij}
u_{ij} \left( {{\bf r},t} \right)_{s{\bf q}}  = \frac{1}{{L^{{3 \mathord{
\left/ {\vphantom {3 2}} \right.
 \kern-\nulldelimiterspace} 2}} }}\left[ {\frac{{\hbar a_0^3 }}{{2M_0
 \omega _{s{\bf q}} }}} \right]^{{1 \mathord{\left/
 {\vphantom {1 2}} \right.
 \kern-\nulldelimiterspace} 2}} \frac{1}{2}i\left( {d_{s{\bf q}}^i q_j  +
 d_{s{\bf q}}^j q_i } \right)e^{i{\bf qr} - i\omega _{s{\bf q}} t}  + c.c.
\end{equation}

Substituting (\ref{u_ij}) into (\ref{w_def}), one  find deformation
potential from one acoustic phonon
\begin{equation}
w\left( {{\bf r},t} \right)_{s{\bf q}}  = \frac{1}{{L^{{3 \mathord{\left/
 {\vphantom {3 2}} \right.
 \kern-\nulldelimiterspace} 2}} }}\left[ {\frac{{\hbar a_0^3 }}{{2M_0
 \omega _{s{\bf q}} }}} \right]^{{1 \mathord{\left/
 {\vphantom {1 2}} \right.
 \kern-\nulldelimiterspace} 2}} e^{i{\bf qr} - i\omega _{s{\bf q}} t}
 \Xi _{s{\bf q}} iq + c.c.,
\end{equation}
where the effective constant of deformation potential of  wave is
included
\begin{equation}
\Xi _{s{\bf q}}  = \Xi ^{ij} \frac{1}{{2q}}\left( {d_{s{\bf q}}^i q_j  +
                   d_{s{\bf q}}^j q_i } \right),
\end{equation}
it depends only on a direction and polarization of  phonon. In our case of
cubic crystal we have $\Xi _{s{\bf q}} = \Xi$ (7\,eV for GaAs
for longitudinal phonons and $\Xi _{s{\bf q}} = 0$
for transverse ones~\cite{Shur}).

So, in Si and GaAs the deformation acoustic phonons only with a longitudinal
polarization can be emitted. The matrix element will finally look like
\begin{equation}
M_{DA}  = i\Xi \sqrt {\frac{\hbar }{{2\rho s}}} L^{ - {3 \mathord{\left/
 {\vphantom {3 2}} \right.
 \kern-\nulldelimiterspace} 2}} \sqrt q I\left( {\bf q} \right) .
\end{equation}
Then probability of deformation acoustic phonon $s{\bf q}$ emission
\begin{equation} \label{w10_def}
w_{10}  = \frac{{\pi \Xi ^2 }}{{\rho sL^3 }}q\left| {I\left( {\bf q} \right)}
 \right|^2 \delta \left( {\varepsilon _{10}  - \hbar sq} \right) .
\end{equation}

Finally, substituting (\ref{w10_def}) into (\ref{W}), passing to spherical
coordinates by the rule (\ref{spher}) and making integration over
delta-function, we have a total probability of generation of
deformation acoustic phonon
\begin{equation}
W_{DA}  = \frac{{\Xi ^2 \varepsilon _{10}^3 }}{{8\pi ^2 \hbar ^4 s^5 }}
\int\limits_0^{2\pi } {d\varphi \int\limits_{ - {\raise0.7ex\hbox{$\pi $}
 \!\mathord{\left/
 {\vphantom {\pi  2}}\right.\kern-\nulldelimiterspace}
\!\lower0.7ex\hbox{$2$}}}^{{\raise0.7ex\hbox{$\pi $} \!\mathord{\left/
 {\vphantom {\pi  2}}\right.\kern-\nulldelimiterspace}
\!\lower0.7ex\hbox{$2$}}} {d\theta \cos \theta \left| {I\left( {\varepsilon
 _{10} \frac{{\cos \theta }}{{\hbar s}},\varphi } \right)} \right|^2 } }
\end{equation}

\subsection{Results and discussion}

The results of calculation of processes of decoherence by structure owing to
spontaneous emission of particles are given in a Fig.~13. As it can be seen
from the graphics the prevailing mechanism of decoherence is the emission of
polarizing acoustic phonons. However, even this process has small probability
at durations of  step appropriate to a GHz range of clock frequencies for
wide structures (wider than 15~nm). The energy splitting versus
distance between electron density maxima  $r$ shown in Fig.~14 is far above
temperature limit (5~mK). It is worth of noting,
that the process of emission of polarizing acoustic phonons is
characteristic for materials such as GaAs, in Si, for example, owing to
symmetry of  lattice they will not be generated. Thus, in Si the prevailing
mechanism of decoherence will be emission of deformation acoustic phonons.

\section{Conclusions}

	New quantum bit is offered on the basis of spatial states of electrons in
symmetrical semi-conductor quantum dots controlled with the help of voltage on
electrodes. The quantum-mechanical calculation of states of the offered qubit
is fulfilled.
The operation of the one-qubit gate which is carrying out unitary transformation,
and not trivial two-qubit CNOT gate are simulated.
The processes of decoherence were investigated. Our study
comprises spontaneous emission of  photon, deformation
acoustic phonon and piezoelectric acoustic phonon.
The work frequencies of the quantum register
being built using offered quantum bits, lie in a range
convenient for electronic control and achieve 1~GHz. The offered
quantum register is scalable and its size is not limited.
The structure reveals sufficient degree of  coherence that
allows by using of the appropriate methods of error correction
to work during unlimited time.
It is necessary to note, that offered quantum bit can be
realized physically at an existing level of cryogenic engineering and
nanoelectronic technology.

\section*{Acknowledgments}

This work has been supported in part by Russian Ministry of Science
and Technologies within framework of program "Advanced Technologies".


\pagebreak

\section*{Figure captions}

\hspace*{\parindent}Fig.~1. Quantum dot qubit.

Fig.~2. Quantum dot potential profile (greyscale, whiter denotes the higher potential).

Fig.~3. Sketch of CNOT gate ($r$ is a separation between electron density maxima;
        $R$ is a separation between qubits' centers)

Fig.~4--7. Wavefunctions of four bottom states of an electron.

Fig.~8. Wavefunction of $\left| 0\right\rangle$ qubit state.

Fig.~9. Wavefunction of $\left| 1\right\rangle$ qubit state.

Fig.~10. NOT gate operation time diagram. $V_g$ is a control gate voltage,
         $\Delta \omega$ is an energy separation between two bottom states.

Fig.~11. NOT gate duration versus distance
between electron density maxima $r$.

Fig.~12. CNOT gate duration versus distance between dots centers $R$ at different $r$.

Fig.~13. Decoherence due to spontaneous emission of photons and acoustic phonons
         versus $r$.

Fig.~14. Energy splitting $\varepsilon_{10}$ versus distance
between electron density maxima $r$.

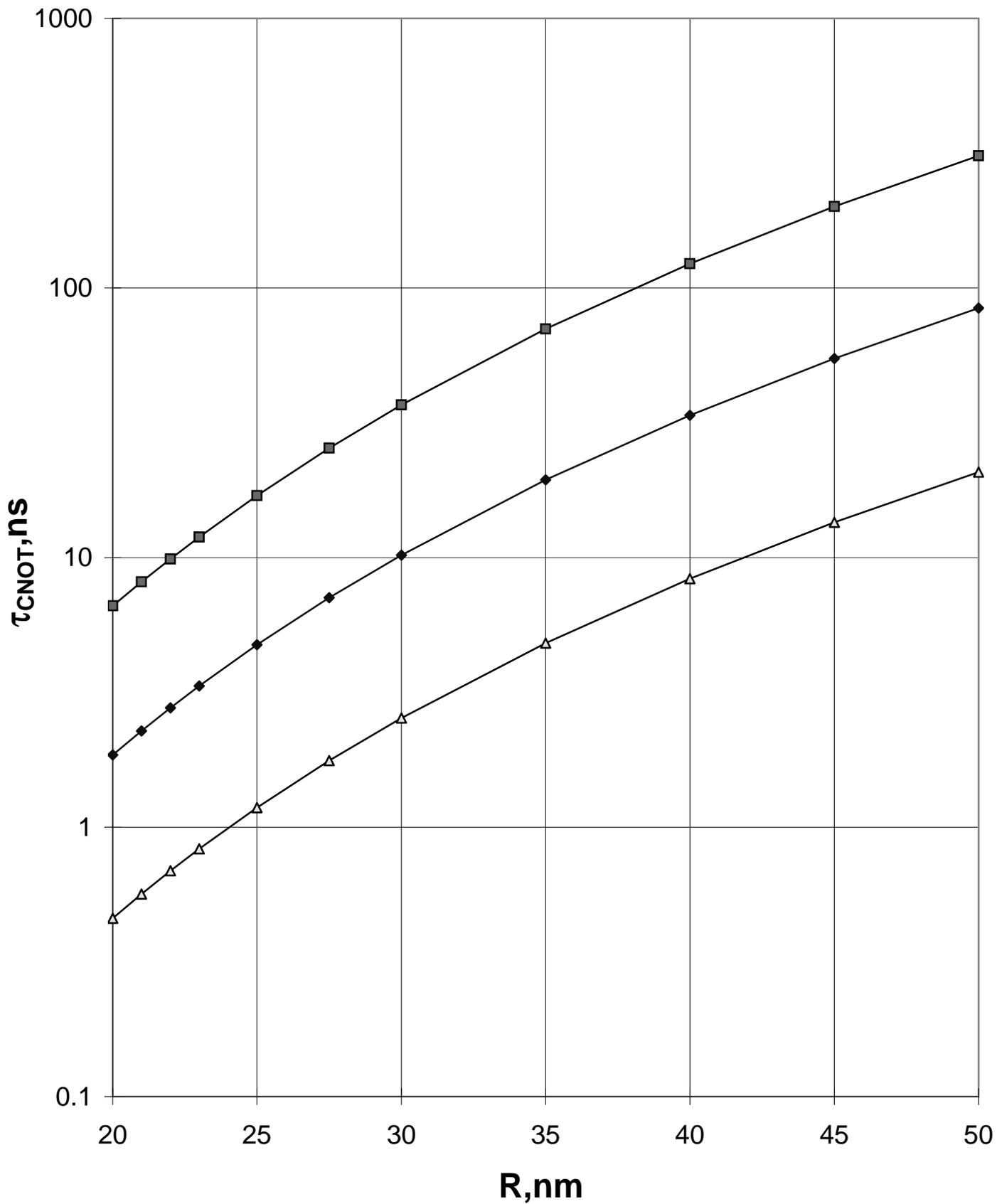
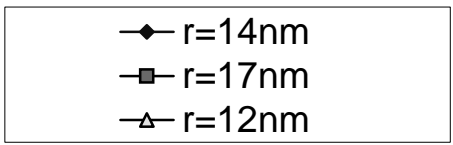
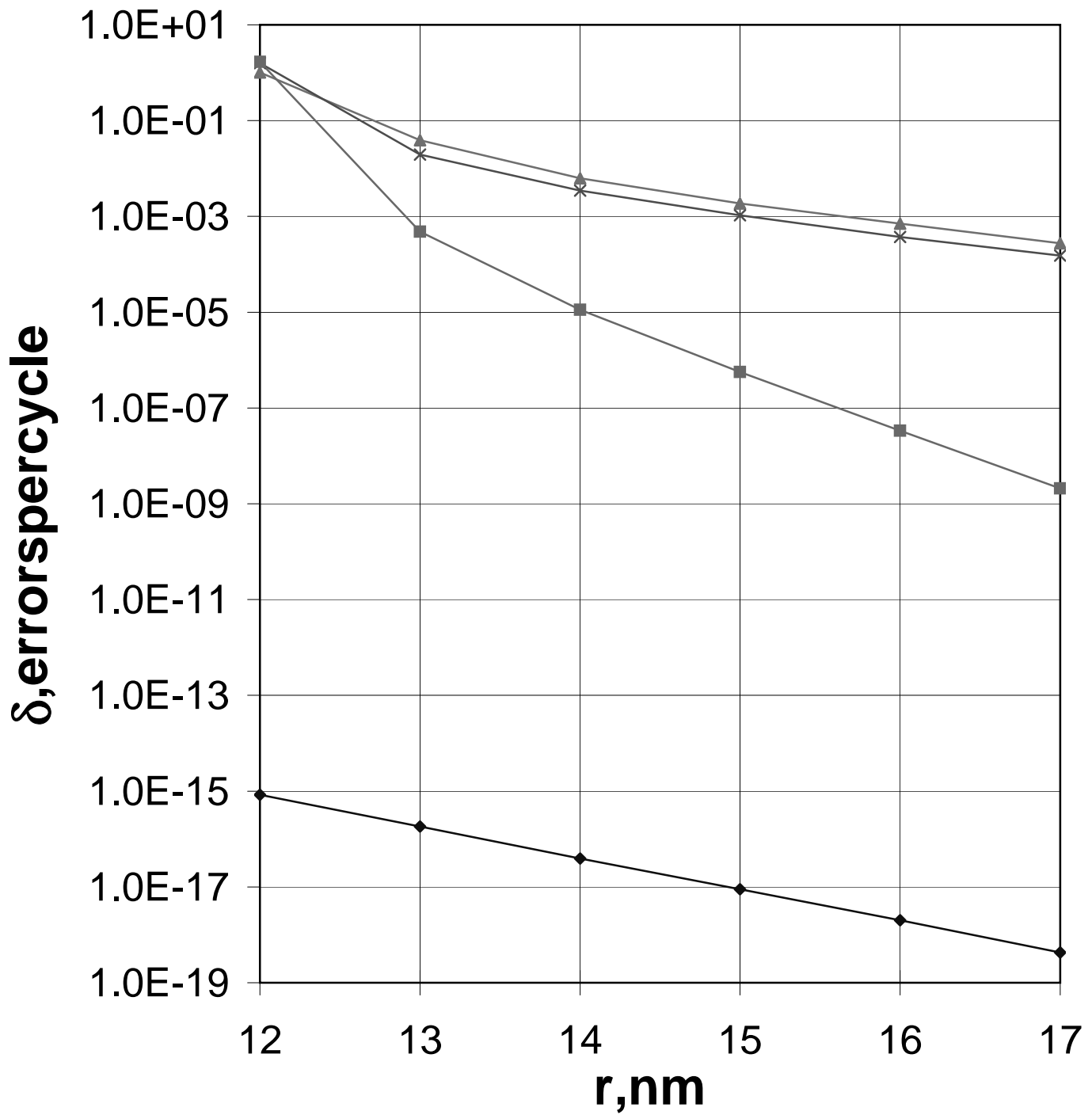
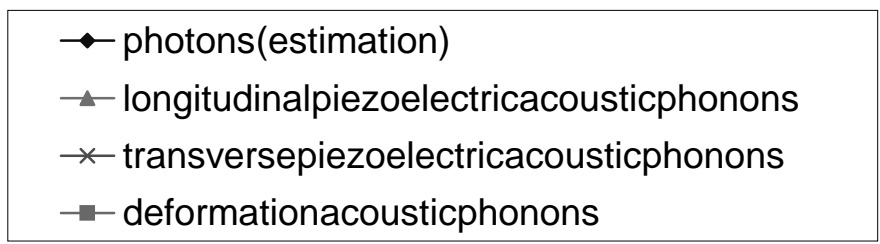
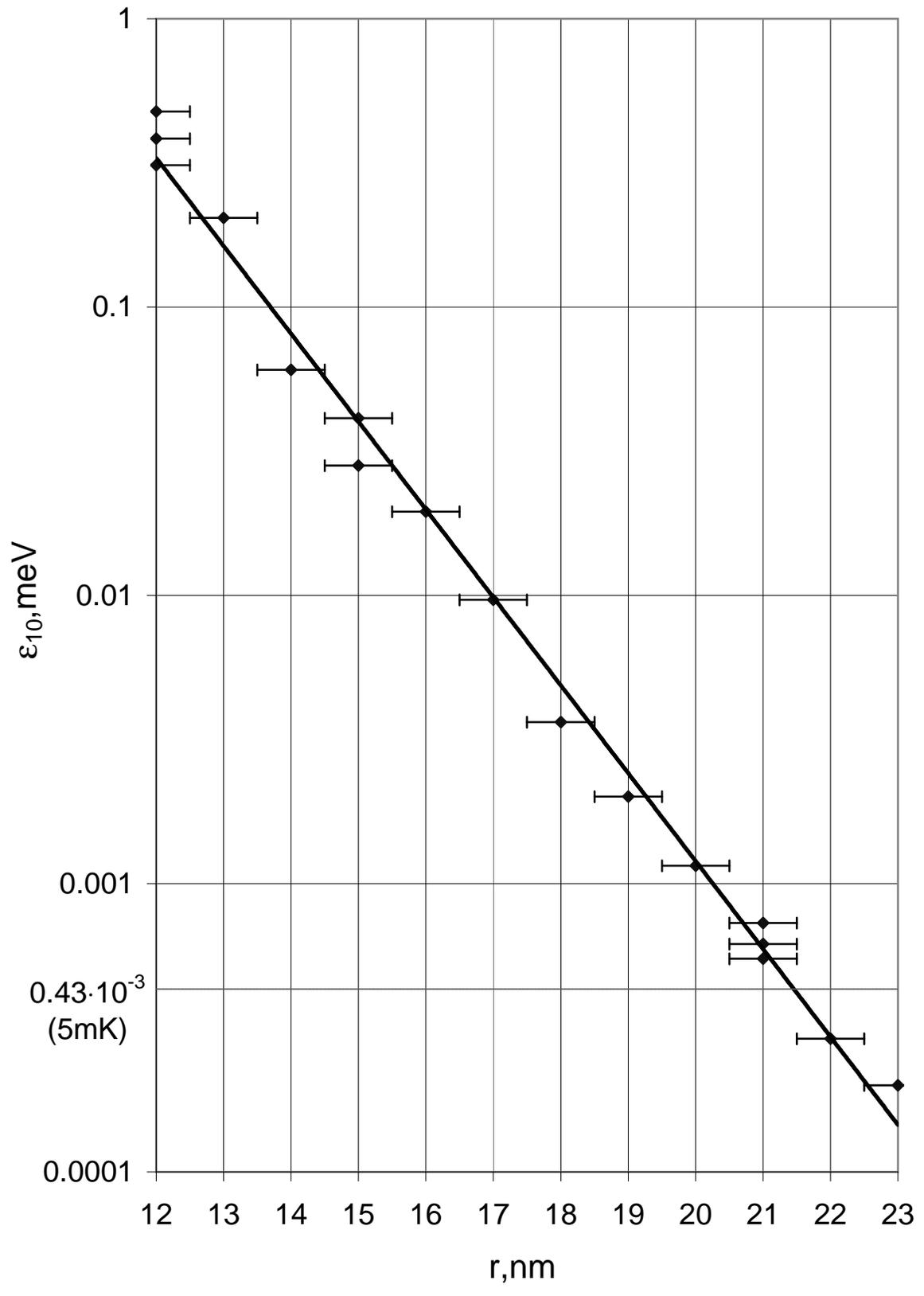
\end{document}